\documentclass[12pt]{article}

\usepackage{graphicx}
\usepackage{color}
\usepackage[colorlinks=true, urlcolor=blue, linkcolor=blue, 
citecolor=blue]{hyperref}
\usepackage{dcolumn}   
\usepackage{bm}        
\usepackage{amssymb}   
\usepackage{cite}

\newcommand{\beq}{\begin{equation}}
\newcommand{\eeq}{\end{equation}}
\newcommand{\beqa}{\begin{eqnarray}}
\newcommand{\eeqa}{\end{eqnarray}}
\newcommand{\beqar}{\begin{eqnarray*}}
\newcommand{\eeqar}{\end{eqnarray*}}

\begin{tiny}

\end{tiny} 

\begin{document}
\thispagestyle{empty}


\vspace{32pt}

\begin{center}

\textbf{\Large A new observable in extensive air showers}

\vspace{50pt}
C.A.~Garc\'\i a Canal$^a$, J.I.~Illana$^b$, M. Masip$^b$, 
S.J.~Sciutto$^a$
\vspace{16pt}

\textit{$^a$IFLP/CONICET and Departamento de F\'\i sica}\\ 
\textit{Universidad Nacional de La Plata, C.C.67, 1900, La Plata, Argentina}\\
\vspace{10pt}
\textit{$^b$CAFPE and Departamento de F{\'\i}sica Te\'orica y del Cosmos}\\
\textit{Universidad de Granada, E-18071 Granada, Spain}\\
\vspace{16pt}

\texttt{cgarciacanal@fisica.unlp.edu.arg, jillana@ugr.es, masip@ugr.es, 
sciutto@fisica.unlp.edu.arg}

\end{center}

\vspace{30pt}

\date{\today}

\begin{abstract}
We find that the ratio $r_{\mu e}$ of the muon to the 
electromagnetic component of an extended air shower at the 
ground level provides an indirect measure of the depth $X_{\rm max}$ 
of the shower maximum. This result, obtained with the air-shower 
code AIRES, is independent of the hadronic model
used in the simulation. We show that the value of
$r_{\mu e}$ in a particular shower discriminates its proton
or iron nature with a 98\% efficiency.
We also show that the eventual production of {\it forward} 
heavy quarks inside the shower 
may introduce anomalous values of $r_{\mu e}$ in isolated 
events.
\end{abstract}

\newpage

\section{ Introduction}
Ultrahigh energy comic rays (CRs) enter the atmosphere 
with energies above $10^{9}\; {\rm GeV} = 1\; {\rm EeV}$. 
The precise determination of their composition, 
direction of arrival and energy provides valuable 
information about their astrophysical sources and about 
the medium that they have traveled through on
their way to the Earth. In addition, their collisions 
with air nuclei probe QCD in a regime never 
tested at colliders. The center of mass energy $\sqrt{2Em_N}$
when the primary CR or the leading hadron inside an 
extensive air shower (EAS)  
hits an atmospheric nucleon is $14$ TeV for $E=10^8$ GeV, 
the nominal energy at the LHC.
Beyond that point collisions occur in uncharted 
territory.

The complementarity between air-shower and collider 
observations does not refer only to the energy involved 
in the collisions, but also to the kinematic 
regions that are accessible in each
type of experiments. At colliders the detectors capable 
of particle identification do not cover the ultraforward 
region, too close to the beampipe. 
This region includes the {\it spectator} 
degrees of freedom in the projectile, which carry 
a large fraction of the incident energy after the collision. 
It turns out that the details there can be relevant to
the longitudinal development of EASs. The
production of forward heavy hadrons \cite{Brodsky:1980pb}, for example, 
is a possibility frequently entertained in the literature
that is difficult to test at colliders \cite{Lykasov:2012hf}.

Air-shower observatories with 
surface detectors able to separate the muon from the 
electromagnetic (EM)
signals, like the Pierre Auger Observatory \cite{AugerOriginal}
will after its projected upgrade \cite{AugerUpgrade}, offer new
oportunities in the characterization of EASs. In this paper we
show that the ratio 
of these two signals at the ground level 
defines a model-independent observable very strongly correlated with the 
atmospheric slant depth of the 
shower maximum and sensitive to possible anomalies introduced 
by forward heavy quarks.

\section{Muons versus electrons in the atmosphere}
An EAS can be understood as the addition of a very 
energetic ({\it leading}) baryon defining the core of 
the shower plus lower energy pions produced
in each collision of this baryon in the air. After just
four interaction lengths (around 300 g/cm$^2$) 99\% of 
the initial energy has already been transferred to pions. 
Neutral pions will decay almost instantly into 
photon pairs, generating the EM
component of the shower, whereas most charged pions 
of $E_{\pi^\pm}\ge 100$ GeV 
will hit an air nucleus giving softer pions. 
Although in hadronic collisions the three pion species 
are created with similar frequency, 
the high-energy $\pi^\pm$s are a source of $\pi^0$s but not the
other way around. As 
a result, 
most of the energy in the EAS will be processed 
through photons and
electrons instead of muons and neutrinos.

At large atmospheric depths the number and the spectrum of
each component in the shower are determined by its very different 
propagation through the air. While electrons and photons basically double 
their number and halve their energy every 37 g/cm$^2$, 
muons lose just a small fraction of energy 
through ionization, bremsstrahlung and pair production 
as they cross the whole atmosphere. Most muons 
created with $E_\mu> 3$ GeV inside the EAS reach the ground.
As a consequence, at the depth $X_{\rm max}$ of the
shower maximum 
electrons dominate over muons 100 to 1, but in inclined 
showers of zenith angle $\theta \ge 60^\circ$ the dominant
signal at the ground level is provided by muons. 
In order to understand this signal, two observations 
are in order.

\begin{enumerate}
\item In inclined events
the EM component at the ground level does not go to zero. 
Although any EM energy deposition high in the atmosphere
will be exponentially
attenuated by the air, there is a continuous production of
photons by high-energy muons: muons do not come alone but 
together with an EM {\it cloud} that is proportional to their number.
\item While the position of $X_{\rm max}$
is dictated by the inelasticity in the first interactions
of the leading hadron and can vary by 
200 g/cm$^2$ among events with identical primaries, 
we expect that the evolution 
beyond the shower maximum is much less {\it fluctuating}. In
particular, the ratio of the muon to the EM component should
depend very mildly on the energy or the nature of the CR primary.
\end{enumerate}
\begin{figure}[!t]
\begin{center}
\includegraphics[width=0.75\linewidth]{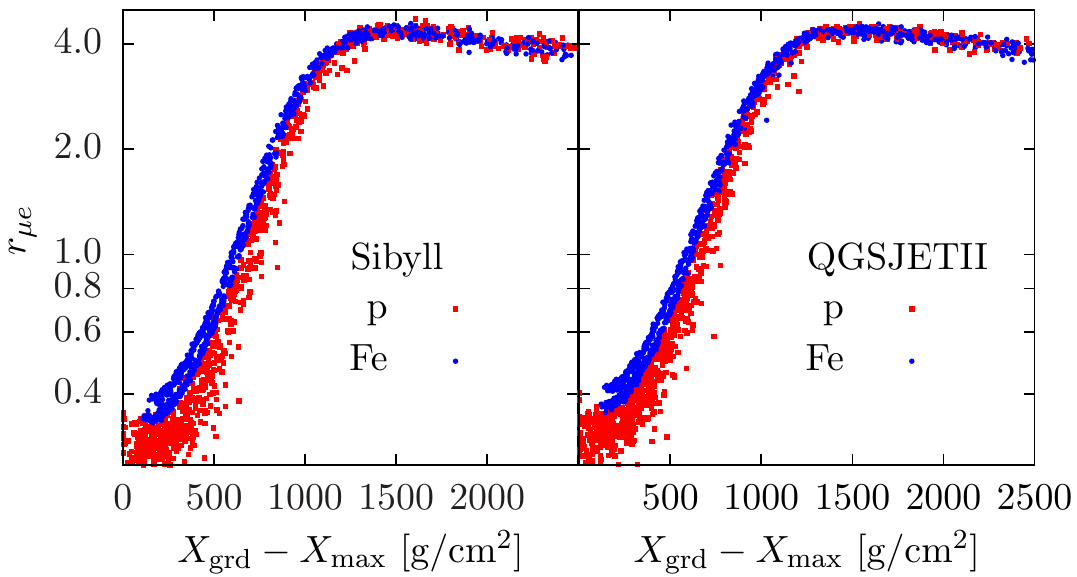}
\end{center}
\caption{$r_{\mu e}={n_\mu\over E_{\rm em}/(0.5\;{\rm GeV})}$ 
versus $X_{\rm grd}-X_{\rm max}$ for proton and
iron showers of 10 and 50 EeV (500 events of each type) simulated
with AIRES using SIBYLL21 (left) and QGSjetII-04 (right). The ground
is at 1400 m of altitude, and we have taken only the particles at 
transverse distances larger than 200 m from the shower axis.
\label{fig1}}
\end{figure}

Fig.~1 fully confirms these two points. We have used the
Monte Carlo code AIRES \cite{AIRES} to simulate 2000 showers of mixed
composition (50\% proton and 50\% iron), different energy 
(50\% 10 EeV and 50\% 50 EeV) and random inclination up to $75^\circ$. 
We have assumed a ground altitude of 1400 m, typical
in EAS observatories. The minimum kinetic energy of
muons, electrons and photons in our simulation is $70$ MeV,
$90$ keV and $90$ keV, respectively. In the figure we plot the
ratio $r_{\mu e}$ between the number of muons and the EM energy
(photons plus electrons) divided by 500 MeV at ground level
in terms of the distance (slant depth) from the ground to the
shower maximum, $X_{\rm grd}-X_{\rm max}$. In our analysis we do 
not include the particles at transverse distances 
from the shower core less than 200 m,
as they tend to saturate the detectors even in inclined events.
The depth $X_{\rm grd}(\theta)$
varies between 800 and 3000 g/cm$^2$ depending on the
inclination of each shower, whereas $X_{\rm max}$ takes typical values
between 700 and 900 g/cm$^2$. We observe that $r_{\mu e}$ is 
a shower observable with relatively small dispersion with the 
energy and the nature of the primary that, for zenith
angles below $60^\circ$, could be used as an indirect measure of 
$X_{\rm max}$. For values between 0.5 and 3 it can be 
approximated by the function
\beq
r_{\mu e} \approx A\, e^{B \,\left( X_{\rm grd}-X_{\rm max} \right)}\,,
\label{relation}
\eeq
whereas at higher inclinations $r_{\mu e}\approx C$ does not
depend on the energy nor the composition
of the CR primary.
In Fig.~\ref{fig1} 
we have used the hadronic models SIBYLL21 \cite{Ahn:2009wx} 
and QGSjetII-04 \cite{Ostapchenko:2013pia}; 
it is most remarkable that this observable is clearly independent from
the hadronic model that we used in the simulation.

\begin{figure}[!t]
\begin{center}
\includegraphics[width=0.7\linewidth]{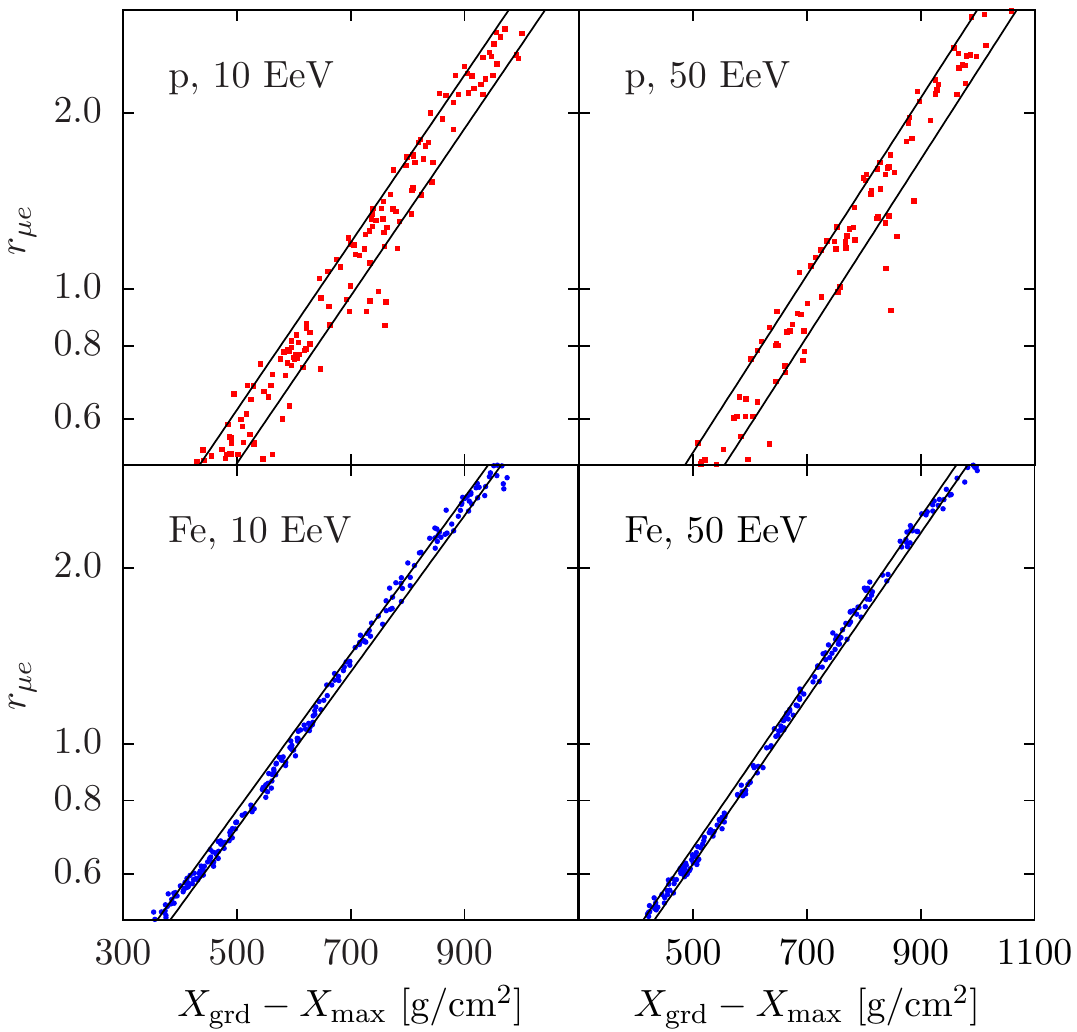}
\end{center}
\caption{Correlation between $r_{\mu e}$ and 
$X_{\rm grd}-X_{\rm max}$ for $0.5 < r_{\mu e} < 3$ and different
CR primaries obtained with SIBYLL21.
\label{fig2}}
\end{figure}

The analysis of the longitudinal development of  EASs 
by a number of authors
\cite{Giller:2004cf,Nerling:2005fj,Schmidt:2007vq,Lipari:2008td,Lafebre:2009en,Andringa:2011zz} shows
that the evolution with the atmospheric depth 
of the EM and the muon components of the shower
can be understood numerically
or with approximate analytical expressions. The average number of
muons and of electrons, however, have large fluctuations
from shower to shower and also a strong dependence on
the hadronic model assumed in each analysis. 
Our result in Figs.~\ref{fig1} and \ref{fig2}
 reflect, basically, that the fluctuations in the
two components of the shower 
are correlated, so that the ratio $r_{\mu e}$
is more stable than the two quantities that define it. 
We will show that this stability can be used to 
discriminate very efficiently the nature of a CR primary.

\section{Composition analyses}
In Fig.~\ref{fig2} we plot the correlation between $r_{\mu e}$ and
$X_{\rm grd}-X_{\rm max}$ for $0.5 < r_{\mu e} < 3$ and 
different primaries. These values of $r_{\mu e}$
include zenith inclinations $33^\circ < \theta < 63^\circ$. 
For example,
a fit with Eq.~(\ref{relation}) for 50 EeV iron primaries gives
(see Fig.~\ref{fig2}) 
\beq
A=0.126 \hspace{1cm} B= 3.25\times 10^{-3} \;{\rm cm}^2/{\rm g}\,,
\eeq
with a dispersion (one standard deviation)
\beq
{\Delta r_{\mu e} \over r_{\mu e}}\approx 0.032\,.
\eeq
The correlation between $r_{\mu e}$ and the shower maximum is then
\beq
X_{\rm max}^{\mu e}=X_{\rm grd}-{\ln \left(r_{\mu e}/A\right) \over B}\pm
{\Delta r_{\mu e}/ r_{\mu e} \over B} \,,
\label{xmax}
\eeq
where the superscript indicates that $X_{\rm max}$ has been
deduced from $r_{\mu e}$ and the uncertainty, around
$10$ g/cm$^2$, corresponds to one standard deviation. Notice that
this uncertainty reflects only the dispersion in the correlation deduced from
our simulation, it does not include the 
experimental error in the determination of $r_{\mu e}$. 
For a 50 EeV proton shower the value of $X_{\rm max}^{\mu e}$ 
obtained this 
way would have a larger uncertainty: our simulation gives
$(A,\, B,\, \Delta r_{\mu e}/ r_{\mu e})=(0.081,\,0.0035\,{\rm cm^2/g},\,0.12)$,
implying a $\pm 34$ g/cm$^2$ dispersion.

Let us discuss with a particular example how 
$r_{\mu e}$ may be used in composition analyses. We simulate a 
50 EeV shower of random inclination and unknown proton or iron 
composition and obtain $r_{\mu e} = 0.648$ and $X_{\rm grd}=1367$ g/cm$^2$
($\theta=50.2^\circ$).
From Eq.~(\ref{xmax}) and this value of $r_{\mu e}$ we know that
if the primary were an iron nucleus 
the shower maximum would be at
$X^{\mu e}_{\rm max}= 863\pm 10$ g/cm$^2$, whereas if it corresponded
to a proton it should be at 
$X^{\mu e}_{\rm max}= 773\pm 34$ g/cm$^2$.
The average values of $X_{\rm max}$ and $\Delta X_{\rm max}$
in 50 EeV showers are
\beqa
{\rm Fe:} &&X_{\rm max}=742\;{\rm g/cm}^2\,,\;\; 
\Delta X_{\rm max}=18\; {\rm g/cm}^2\nonumber \\ 
{\rm H:}  &&X_{\rm max}=838\;{\rm g/cm}^2\,,\;\; 
\Delta X_{\rm max}=52\; {\rm g/cm}^2\,.
\eeqa
Adding the uncertainties in quadrature we see that 
$(863\pm 10) \;{\rm g/cm}^2$ is $5.8\,\sigma$ away from  iron 
[$(742\pm 18)\;{\rm g/cm}^2$], while 
$(773\pm 34)\;{\rm g/cm}^2$ is just $-1.0\,\sigma$ away 
from proton [$(838\pm 52)\;{\rm g/cm}^2$]. 
This clearly reveals the proton nature of the shower. 

The actual value of $X_{\rm max}$ in the 
previous event was $774$ g/cm$^2$. If 
measured with some fluorescence detectors, 
$X_{\rm max}$ would 
also signal the proton nature of the primary: it is 
$1.7\,\sigma$ away from iron and just $-1.2\,\sigma$ from proton.
However, the statistical significance would have
been much lower than the one obtained from $X^{\mu e}_{\rm max}$.
Applying the discriminant deduced from
$r_{\mu e}$ to the 290 events 
in Fig.~\ref{fig2} (50 EeV events with $r_{\mu e}$ between 0.5 and 3)
we find that it gives the right answer in 284 of 
them (98\%), while $X_{\rm max}$ indicates the true proton or
iron nature in 262 events (92\%). 
Notice that two events with similar values of
$r_{\mu e}$ and $(X_{\rm grd} - X_{\rm max})$ may have quite different 
inclination ({\it i.e.}, different $X_{\rm grd}$), especially if 
their composition is different.
As a consequence, the value of 
$X^{\mu e}_{\rm max}$ deduced from $r_{\mu e}$ 
depends on whether the primary is a proton or an iron
nucleus,  {\it separating} both
possibilities from each other further than the direct 
observation of $X_{\rm max}$. Of course,
there could be an experimental error in $r_{\mu e}$ 
(measured at the surface detectors) larger 
than the one in $X_{\rm max}$ (at the fluorescence 
detectors), but the use of this observable 
in composition analyses \cite{Abreu:2013env}
seems very promising. 

Notice also that in our previous analysis we have assumed
a given value for the energy of the EAS. The shower energy 
could in principle 
be deduced from other observables, like the total signal at the
surface detectors, its lateral distribution, etc. If a 
particular observatory is able
to determine $E\pm \Delta E$ with a certain precision, 
then the correlation between $r_{\mu e}$ and $X_{\rm max}$
(the specific values of $A$ and $B$ for this event)
should be established from a fit of 
showers within the same energy interval. As for the 
range of distances to the shower axis 
to be included in the definition of $r_{\mu e}$ (we
have taken all transverse distances beyond  
200 m), the optimal one should be decided after a
simulation of the surface detectors in the particular
observatory.

\section{Forward charm and bottom hadrons}
Our results above show that, while the position of $X_{\rm max}$ 
may have large fluctuations related to the inelasticity in the first 
few interactions of the leading hadron, the longitudinal
evolution of an EAS from that point to the ground 
is very stable, and the ratio $r_{\mu e}$ 
appears always strongly correlated with $X_{\rm grd}-X_{\rm max}$. 
The obvious question would then concern the possibility to break
this correlation: what physical process
could explain an anomalous value of $r_{\mu e}$? 

As we have mentioned before, the production of forward
heavy hadrons carrying a large fraction of the incident 
energy is a possibility often discussed in the literature. 
Analogous processes ($p\to K^+\Lambda$) \cite{Edwards:1978mc}
have been observed for strange particles.
Indeed, the asymmetry 
detected in charm production at large Feynman $x$
\cite{Alves:1996rz} indicates a soft contribution that may
be explained with an {\it intrinsic charm} hypothesis
\cite{Brodsky:1980pb,Lykasov:2012hf} or through
the {\it coalescence} of perturbative charm 
with the valence quarks present in the 
projectile \cite{Barger:1981rx,Vogt:1995fsa} (this has also
been the approach in SIBYLL 2.3 \cite{Engel:2015dxa}).

Charm or bottom hadrons 
produced inside an EAS with energy above $10^{9}$ GeV would be 
long lived (their decay length becomes larger than 100 km) and very
penetrating: a $D$  or a $B$ meson would keep  
$60\%$ \cite{Barcelo:2010xp} or $80\%$ \cite{Bueno:2011nt} of its energy in 
each collision with the air, respectively. One of these mesons 
could experience 10 ($D$)
to 20 ($B$) collisions before its energy has been
reduced to $\approx 10^{7}$ GeV and it decays.
It would be a small fraction of the total energy in the
shower, but if the deposition takes place near the ground 
it may reduce significantly the value of $r_{\mu e}$. This observable
could then open new possibilities in the search for heavy quark effects
in EASs \cite{Bueno:2013rn}.

We have used AIRES \cite{AIRES} for a first look at this 
issue. Although AIRES includes the production of {\it central} 
(perturbative) heavy hadrons as well as their propagation 
in the atmosphere \cite{Canal:2012uh}, we find that 
these hadrons do not carry enough energy to have any influence 
on $r_{\mu e}$. Therefore, we have simulated events where 
the leading hadron may create a forward
charmed or bottom hadron that takes a large fraction 
of its energy (to be definite, we have used the
$x$ distribution in \cite{Paiva:1996dd}).
We have run events with 10 and 50 EeV of energy, 
arbitrary inclination and a proton or 
iron primary (in the second case the heavy
hadron will take a fraction of the energy per nucleon in the
projectile).
Although the average value of $r_{\mu e}$ 
is not changed significantly by the forward
heavy hadrons, we are able to identify two types of 
isolated events that are clearly anomalous.
\begin{itemize}
\item The first anomaly may appear in proton showers
when the leading hadron 
creates a $B$ meson or a $\Lambda_b$ baryon 
of energy above 1 EeV. 
These hadrons are then able to penetrate
very deep in the atmosphere and decay near the ground, 
starting a {\it minishower} of $10^6$--$10^8$ GeV
that reduces the value of $r_{\mu e}$.
The anomaly only appears  
in showers with $50^\circ < \theta < 60^\circ$:
at lower zenith angles the relative effect of the minishower
is too small (the attenuation of the rest of the shower at the 
ground level is insufficient), whereas in showers with a larger 
inclination the heavy hadron tends to decay too far
from the ground. We find events where the actual
$X_{\rm max}$ is 400 g/cm$^2$ smaller than the depth $X^{\mu e}_{\rm max}$
deduced from $r_{\mu e}$, a 12$\sigma$ deviation.
\item The second anomaly is an
indirect effect of the heavy quarks: it appears in
very inclined EASs when a muon of $E_\mu\ge 10^7$ GeV 
experiences a relatively hard radiative process
(bremsstrahlung or pair production) near the ground.
At such high energies pions and kaons are very long lived,
and the main source of muons is the 
decay of charm and bottom hadrons (see \cite{Illana:2010gh} for 
other sources of atmospheric muons).
We find that the effect may only appear at 
zenith angles $\theta > 65^\circ$. These inclinations favor the
decay of the heavy hadrons high in the atmosphere, 
before they lose energy. We identify events where a high-energy
muon crosses $2000$--$3000$ g/cm$^2$ of air
and deposits $10^6$--$10^7$ GeV at $100$--$500$ g/cm$^2$ from
the ground, changing the muon-to-EM ratio $r_{\mu e}$ from the 
asymptotic value
$C\approx 4$ to a value around $1$. Since the muon
comes from a forward heavy hadron, in these events the anomaly 
is larger near the shower core, and it disappears as we increase the 
lateral distance.
\end{itemize}

\section{Summary and discussion}
The possibility to separate  the muon and the EM components 
in the surface detectors at CR observatories
seems essential both to fully characterize the
shower and also to {\it tune} the Monte Carlo codes
used to simulate  ultrahigh-energy events. 
Here we have discussed
a new observable, the ratio $r_{\mu e}$ between the two 
components, that correlates with
$X_{\rm max}$ with an uncertainty of around 
$\pm 10$ g/cm$^2$ for iron
nuclei or $\pm 40$ g/cm$^2$ for protons. A precise analysis of the 
spectrum and the composition of 
ultrahigh energy CRs relies very strongly on
simulations, and this observable could provide a crucial 
consistency check. 
In particular, it could give
a surprisingly effective discriminant in composition
analyses.

One important issue currently being discussed \cite{Aab:2015bza}
is the possible under-prediction of the muon signal by
basically all hadronic simulators. This would suggest a
correction towards 
a higher multiplicity in hadron collisions: 
a larger number
of less energetic pions inside the shower implies a stronger
muon signal (number of muons) with the same EM signal (energy
in electrons and photons). Obviously, if the {\it muon problem} 
is confirmed after the upgrade of the Auger observatory 
and the hadronic 
models are modified, their prediction for $r_{\mu e}$ will
change accordingly. The analysis with the {\it wrong} 
simulators presented here would then be biased, and our
determination of $X_{\rm max}$ from $r_{\mu e}$ would have a
{\it systematic} error. The only way to identify and correct
this bias would be to compare $X^{\mu e}_{\rm max}$ with the
$X_{\rm max}$ provided by the fluorescence detectors in hybrid
events. It is then interesting that such comparison can
be used to quantify the suspected muon problem of
current simulations.

Our analyses based on SIBYLL and QGSjetII 
show that the relation between $X_{\rm max}$ 
and $r_{\mu e}$ is very stable and model independent. 
It is crucial that we 
compare showers at the same distance depth from the
maximum ({\it i.e.}, same value of $X_{\rm grd}-X_{\rm max}$),
which minimizes the shower to shower fluctuations. 
Our results also reflect that the fluctuations 
and the model dependencies in the muon and the EM components
of a shower are correlated, {\it i.e.}, if $r_{\mu e} = x/y$
with $x=n_\mu$ and $y=E_{\rm em}/(0.5\;{\rm GeV})$, then
$\Delta r_{\mu e} \ll \sqrt{ \left( \Delta x/y \right)^2
+ \left( \Delta y\;x/y^2 \right)^2 }$.

We have argued that only the production of
very energetic {\it forward} heavy hadrons could introduce 
anomalies. In particular, we have identified reductions in 
the value of 
$r_{\mu e}$ caused {\it (i)} by the decay of these hadrons deep in the
atmosphere in proton showers of intermediate 
inclination ($50^\circ < \theta <60^\circ$), and {\it (ii)} by 
stochastic energy depositions near the ground coming from very 
energetic muons in inclined showers ($\theta > 65^\circ$).
These muons would be created high in 
the atmosphere through semileptonic decays of charm and bottom hadrons.
Therefore, we conclude that  
$r_{\mu e}$ may be a key observable to characterize EASs, determine
the nature of the CR primary, 
and even
in the search for the elusive forward heavy hadrons.

\section*{Acknowledgments}
We would like to thank Antonio Bueno and Carlos Hojvat for 
enlightening discussions and Stephane Coutu for a careful reading
of the manuscript.
This work has been supported by 
ANPCyT and CONICET of Argentina, by MICINN of Spain (FPA2013-47836,
FPA2015-68783-REDT and  
Consolider-Ingenio {\bf Multidark} CSD2009-00064) and by Junta de 
Andaluc\'\i a (FQM101).

\end{document}